\documentclass[aps,prl,superscriptaddress,twocolumn,a4paper,showpacs]{revtex4}
\usepackage{graphicx}
\usepackage{bm}
\usepackage[dvips]{color}

\begin{document}
\title{Measurement of the Proton Asymmetry Parameter $\bm{C}$ in Neutron Beta Decay}
\date{\today}

\author{M.~Schumann}
\email{marc.schumann@gmx.net}
\affiliation{Physikalisches Institut der Universit\"at Heidelberg, Philosophenweg 12, 69120 Heidelberg, Germany}
\author{M.~Kreuz}
\affiliation{Institut Laue-Langevin, B.P. 156, 38042 Grenoble Cedex 9, France}
\author{M.~Deissenroth}
\affiliation{Physikalisches Institut der Universit\"at Heidelberg, Philosophenweg 12, 69120 Heidelberg, Germany}
\author{F.~Gl\"{u}ck}
\affiliation{IKEP, Universit\"at Karlsruhe (TH), Kaiserstr.\ 12, 76131 Karlsruhe, Germany}
\affiliation{Research Institute for Nuclear and Particle Physics, POB 49, 1525 Budapest, Hungary}
\author{J.~Krempel}
\altaffiliation{present address: Institut Laue-Langevin, Grenoble}
\affiliation{Physikalisches Institut der Universit\"at Heidelberg, Philosophenweg 12, 69120 Heidelberg, Germany}
\author{B.~M\"{a}rkisch}
\affiliation{Physikalisches Institut der Universit\"at Heidelberg, Philosophenweg 12, 69120 Heidelberg, Germany}
\author{D.~Mund}
\affiliation{Physikalisches Institut der Universit\"at Heidelberg, Philosophenweg 12, 69120 Heidelberg, Germany}
\author{A.~Petoukhov}
\affiliation{Institut Laue-Langevin, B.P. 156, 38042 Grenoble Cedex 9, France}
\author{T.~Soldner}
\affiliation{Institut Laue-Langevin, B.P. 156, 38042 Grenoble Cedex 9, France}
\author{H.~Abele}
\email{abele@e18.physik.tu-muenchen.de}
\affiliation{Physikalisches Institut der Universit\"at Heidelberg, Philosophenweg 12, 69120 Heidelberg, Germany}

\begin{abstract}
The proton asymmetry parameter $C$ in neutron decay describes the correlation between neutron spin and
proton momentum. In this Letter, the first measurement of this quantity is presented. The result
$C=-0.2377(26)$ agrees with the Standard Model expectation. The  
coefficient $C$ provides an additional parameter for new and improved Standard Model tests.
From a differential analysis of the same data (assuming the Standard Model), we obtain 
$\lambda=-1.275(16)$ as ratio of axial-vector
and vector coupling constant.
\end{abstract}


\maketitle

The free neutron, decaying into electron, proton, and 
anti-electron-neutrino (in the following called neutrino),
constitutes a well suited low energy laboratory to study
the structure of semileptonic weak interactions 
and to probe the underlying symmetries. It is
a rather simple system without any corrections due to nuclear structure. 
In the $V\!-\!A$ description of the Standard Model (SM), it is governed by two free 
parameters only (the Fermi constant
$G\!_F$ is considered to be known exactly and CP violation can be neglected): The first element
of the quark mixing matrix, $V_{ud}$, and the ratio of axial-vector
and vector coupling constant, $\lambda=g_A/g_V$. On the other hand, there are much more 
experimentally accessible
observables, such as the lifetime $\tau_n$ and various correlations
between spins and momenta of the neutron and the decay products. Thus, the problem
is overdetermined and permits to test the SM description of neutron
decay in many ways \cite{Abe07,Sev06}.

Prominent examples for angular correlation coefficients
are the electron asymmetry parameter $A$ and the neutrino asymmetry parameter $B$,
correlating neutron spin with electron 
and neutrino momentum, respectively, and the correlation $a$ between
the momenta of electron and neutrino. Within the SM, each of these parameters
allows to derive $\lambda$. They
have been measured several times with increasing precision and are
still in the focus of current research. However, 
the correlation between neutron spin
and proton momentum, the proton asymmetry parameter $C$, has not been determined experimentally so far.
It describes the proton angular distribution \cite{Tre58}
\begin{equation}
\omega(\theta) \ \text{d}\theta = ( 1 + C \ \cos \theta ) \ \text{d}\theta 
\end{equation} 
with the angle $\theta$ between
neutron spin and proton momentum. This Letter reports on the first measurements of $C$.

A precisely known parameter $C$ permits new cross-checks within the SM
at low energies as the three decay products are
kinematically coupled,
\begin{equation}\label{eq_relat}
C = - x_C(A+B)\textnormal{,}
\end{equation}
where $x_C=0.27484$ is a kinematical factor \cite{Tre58,Glu96}. 
Within the SM, $C$ gives an experimentally different access to $\lambda$
($g_V$ and $g_A$ are assumed to be real here):
\begin{equation}\label{eq_CandLambda}
C = x_C \frac{4 \ \lambda}{1+3\lambda^2} \textnormal{,}
\end{equation}
although with somewhat smaller sensitivity than $a$ or $A$.
Physics beyond the SM, e.g. $V+A$-, scalar- or tensor-like interactions,
may alter this relation \cite{Glu96,Sju05}. 
Thus, $C$ is a new input parameter for global analyses
of neutron decay to constrain the size of possible ``new physics'' effects.

A first measurement of the proton asymmetry parameter $C$ using the electron
spectrometer PERKEO II has been carried out in 2001 \cite{Kre04}. In the
following, we report on the succeeding experiment of 2004, performed
with much cleaner systematics and improved statistics.

PERKEO II was installed at the cold neutron beam 
position PF1B \cite{abele2006} of the Institut Laue-Langevin (ILL), Grenoble. Details
on the experimental setup can be found in \cite{Sch07,Sch07b}. The neutron beam
was spin polarized using two super mirror polarizers in X-SM geometry \cite{kreuz2005}.
Polarization $P=0.997(1)$ and spin-flipper efficiency $F=1.000(1)$
were independent of position and wavelength, and stayed constant during the measurement.

PERKEO II  
consists of a pair of superconducting coils in split pair configuration, generating
a magnetic field perpendicular to the neutron beam, with a maximum of 1.03 T.
The neutron spins
align parallel to the field, that thereby divides the full solid
angle into one hemisphere in and one against spin direction. Each hemisphere was
covered by an electron detector consisting of a large area plastic scintillator with
photomultiplier readout. All charged decay products generated 
in the spectrometer center (``decay volume'') were guided onto one of the
detectors by the magnetic field, realizing full $2 \times 2 \pi$ coverage.

Protons were detected in coincidence with the decay electrons alike in refs.\ \cite{kreuz2005B,Sch07}. 
The typical energies of electron and proton differ by three orders of magnitude. In order to
detect both particles with the same detector, 
the protons were accelerated onto a thin (\mbox{15$-$30 $\mu$g cm$^{-2}$})
carbon foil on negative high voltage (HV). Thereby they gained enough energy to release one
or several secondary electrons from the foil, which were accelerated to the
plastic scintillator on ground. The decay volume itself was kept on ground
potential by four grounded grids made from AlSi wires on each side. 
Thus all protons gained the same
acceleration and the assignment of the proton to the hemisphere was not affected by the
accelerating HV. The detection of the primary decay electron was not altered by
the HV if the initial kinetic energy was above 84 keV, regardless of the
emission angle.
The energy loss of the electrons in the carbon
foil was very small ($\sim$0.5 keV) and constant for all relevant electron energies, thus 
not leading to a systematic effect. The detection signature
was an electron trigger, whose energy was recorded, followed
by a proton signal. The latter arrived delayed due to the 
finite drift time from the decay vertex
to the region of acceleration. No information on the proton
energy was available with this method as the proton was detected after acceleration and as
its drift time in the grounded region strongly depended on the position of the decay
vertex and the emission angle relative to the magnetic field lines.

For each detector, we obtain four spectra,
defined by particle emission in the hemisphere in ($+$) or against ($-$) neutron spin direction:
$Q^{++}(E)$, $Q^{-+}(E)$, $Q^{--}(E)$, and $Q^{+-}(E)$.
The first sign denotes the electron, the second the proton, and $E$ 
the electron's energy. In order to derive the proton asymmetry, we have to
integrate out the electron (energy and emission direction). With
\begin{equation}\label{eq_integrals}
\begin{array}{rcl}
\rho^+&=&\int (Q^{++}(E)+Q^{-+}(E)) \ \text{d}E\qquad\text{and} \vspace{0.1cm} \\
\rho^-&=&\int (Q^{+-}(E)+Q^{--}(E)) \ \text{d}E
\end{array}
\end{equation}
we denote the number of protons that were emitted into one hemisphere with the neutron
spin pointing towards and away from this hemisphere, respectively. 
In this notation, the proton asymmetry is
\begin{equation}\label{eq_proton}
  C = \frac{\rho^+ -\rho^-}{\rho^+ +\rho^-}.
\end{equation}
Note that we use only one proton hemisphere and flip the spin, thus avoiding
systematic effects due to different proton efficiencies.
The definition (\ref{eq_proton}) has an opposite sign compared to \cite{Glu95,Glu96} to
maintain the convention that a positive asymmetry parameter indicates more particles to be
emitted in spin direction.

The finite electron detector threshold, the HV-barrier modifying the
electron detection at low energies, and electron backscattering effects
prevent the evaluation of the integrals (\ref{eq_integrals}) in the whole energy range.
Furthermore, the integrals combine events with electron detection in two different
detectors, without cancellation of the related detector efficiencies.
Therefore the measured $Q$-spectra were fit by theoretical functions in regions high
enough to describe the spectra properly and where the detector efficiencies for
electrons were 1. The spectra were then extrapolated to the full energy range and
integrated, taking into account the finite detector energy resolution.

The theoretical functions to describe the $Q$-spectra are given in ref \cite{Glu95}
with $Q^{ij}\!=\!q^{ij} F(E)$ and the Fermi function $F(E)$:
\begin{eqnarray} \label{QFunktion}
q^{++}_{r<1} \! &=& \begin{array}{l} \!\frac{2\!-\!r}{2} \!+\! \frac{a \beta }{4} \left(\frac{r^2}{2}\!-\!1 \right) \!+\! \frac{P A \beta}{2} \left(1\!-\!\frac {2r}{3} \right) \!+\! \frac{PB}{2} \left(\frac{r^2}{3}\!-\!1 \right)
\end{array}
 \nonumber \\
q^{++}_{r \ge 1} \! &=& \begin{array}{l} \!\frac{1}{2r} \ \left( 1 - \frac{a \beta}{4 r} + \frac{P A \beta}{3 r} -\frac{2 P B}{3} \right) \end{array} \nonumber \\
q^{--}\!&=&\! q^{++}[P \to -P] \\
q^{+-}\!&=&\! 2+PA \beta - q^{++} \nonumber \\ 
q^{-+}\!&=&\! 2-PA \beta - q^{--} \qquad \text{with} \nonumber \\
r &=& \beta \begin{array}{l} \frac{E+m_{\text{e}}}{E_{\text{max}}-E} \end{array} \text{.}\nonumber
\end{eqnarray}
$A$ and $B$ are the electron and the neutrino asymmetry parameter,
respectively. $P$ is the neutron beam polarization, 
$\beta=v/c$. $E_{\text{max}}\!=\!782$ keV is the endpoint of the beta spectrum, 
$m_{\text{e}}$ the electron mass.
Definition (\ref{QFunktion}) is separated into two regions ($r<1$, $r \ge 1$) by the energy dependent parameter
$r$, with $r=1$ for $E=236$ keV.

\begin{figure*}[tb]
\begin{minipage}{8.6cm}
\begin{center}
\includegraphics*[width=8.5cm]{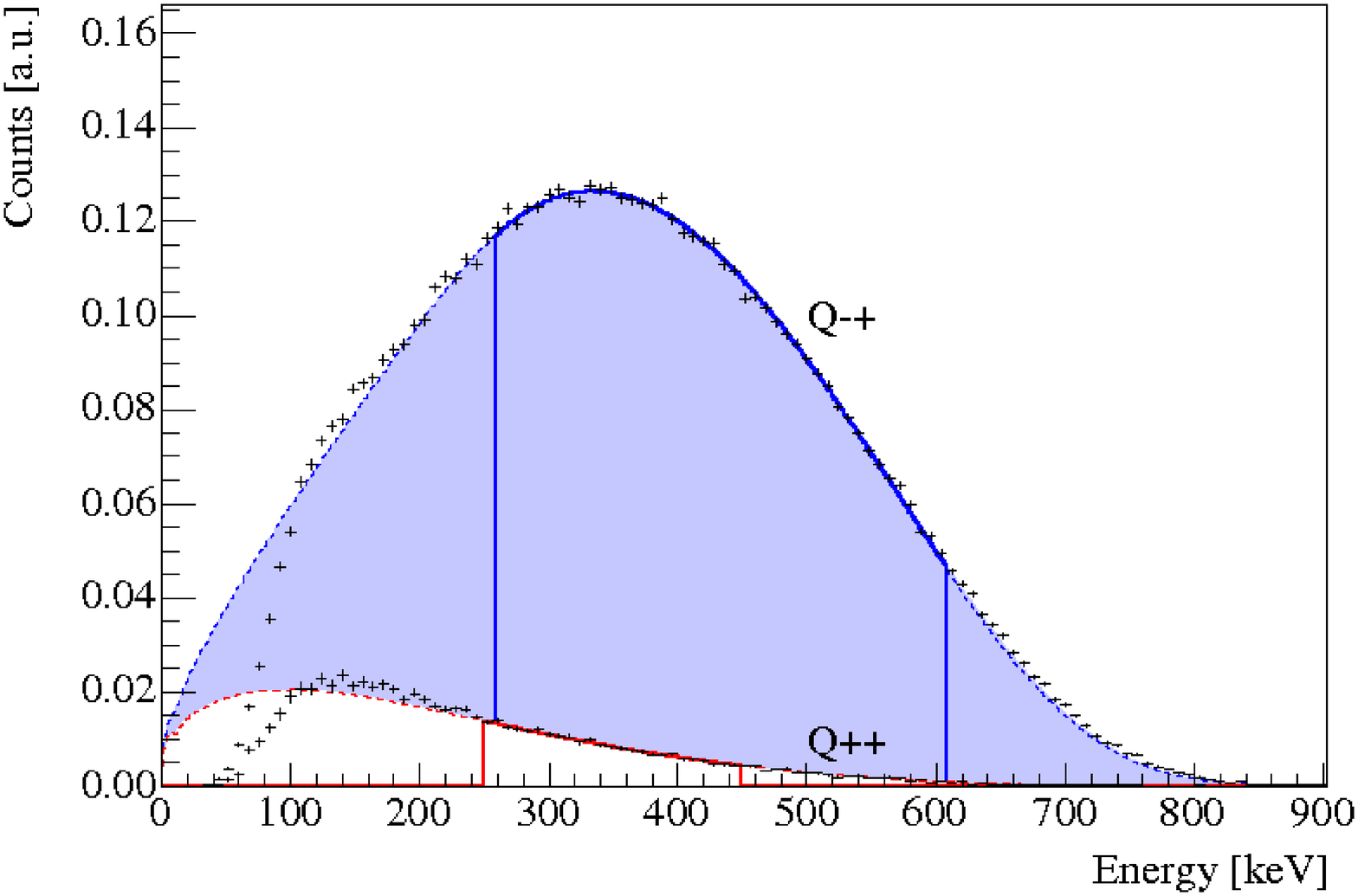}
\end{center}
\end{minipage}
\hfill
\begin{minipage}{8.6cm}
\begin{center}
\includegraphics*[width=8.5cm]{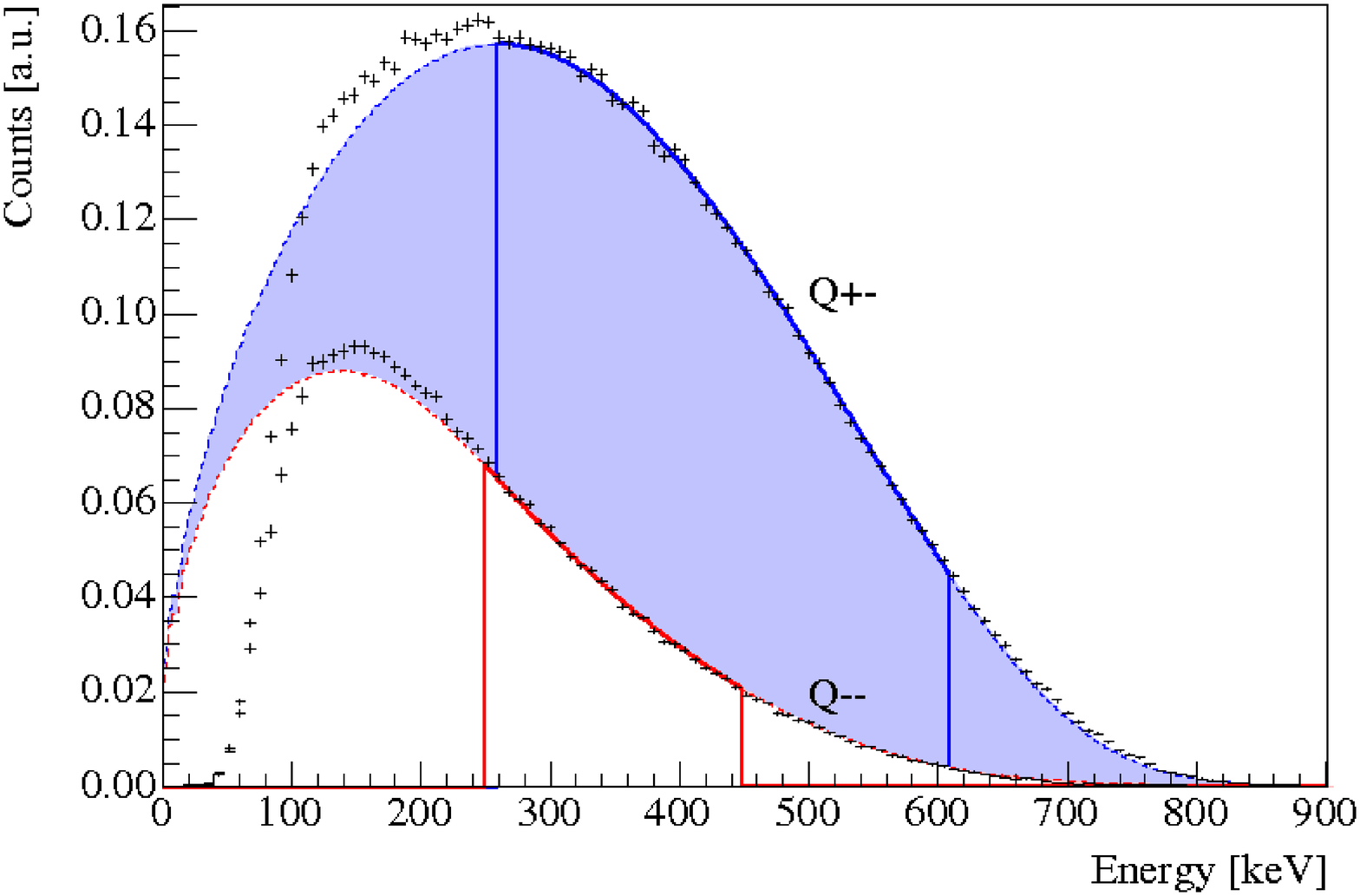}
\end{center}
\end{minipage}
\caption{\label{Fig_CFits}The spectra to determine $C$ from proton detector 2
were fitted in the regions indicated
by the solid lines. The functions were extrapolated to cover the whole
energy range and finally integrated (filled areas).
The one parameter fits describe the spectra very well
for energies above 240 keV. The
agreement between fit and data in the low energy regime can be improved by
including the detector threshold and a backscattering correction to the
theoretical description. We refrain from this approach since the
uncorrected $Q$-spectra, as shown in this figure, are needed to determine $C$.
Since the magnetic mirror effect (cf.\ text
and \cite{Sch07}) acts
differently on $Q^{ii}$ (electron and proton in same hemisphere) and
$Q^{ij},~i\ne j$ (electron and proton in different hemispheres),
the fit regions are different.}
\end{figure*}

The electron detectors were calibrated using six monoenergetic
conversion electron peaks ($^{109}$Cd, $^{139}$Ce, $^{113}$Sn, $^{137}$Cs, and two peaks 
of $^{207}$Bi) covering the relevant energy range up to 1 MeV. They were
inserted into the spectrometer center regularly. To correct for spatial detector effects, the
whole decay volume projection was scanned two-dimensionally several times. Above \mbox{100 keV}, 
detector response was linear within \mbox{0.8\%}. At lower energies, 
the linear relation $E = g k + E_{\text{off}}$
between energy $E$ and analog-to-digital converter (ADC) channel $k$ 
was no longer valid. The detector gain $g$ was known
with a precision of 0.3\% (0.6\%) for detector 1 (detector 2). The energy
offsets of the two detectors were $E_{\text{off,1}}=(37.7\pm1.0)~\text{keV}$ and
$E_{\text{off,2}}=(40.4\pm2.1)~\text{keV}$, respectively. 
Their uncertainty is the limiting factor for the measurement of $C$ as it 
directly enters the $Q^{ij}$ fit functions and thus alters extrapolation and integration 
results. It leads to an uncertainty of 0.82\% on $C$. The uncertainties on
detector gain and the energy resolution of about \mbox{15\%} (FWHM) at 
\mbox{1 MeV} lead to much lower errors (\mbox{0.38\%} and \mbox{0.12\%}, respectively).

A valid event consisted of an electron trigger followed by
a coincident delayed proton signal. All delayed signals within a window of \mbox{82 $\mu$s}
were recorded for both detectors. The first 40 $\mu$s defined the coincidence 
window. It was chosen such
that the number of protons with higher flight-times was almost negligible (0.03(1)\% correction on $C$). Accidental coincidences were measured in the second half of the \mbox{82 $\mu$s} window 
(delayed coincidence method). 
The correction of $-0.81(15)$\% due to accidental coincidences was obtained
directly from measured data as the timing information for all delayed signals was available for all events.

The fit regions to evaluate the $Q$ spectra were chosen according to the detailed analysis 
of the neutrino asymmetry parameters \cite{Sch07,Sch07b} generated from same 
($Q^{++}$, $Q^{--}$) and opposite hemisphere ($Q^{+-}$, $Q^{-+}$) spectra.
Due to the finite neutron beam width, some particles were
emitted towards an increasing magnetic field and could be reflected (depending on their emission angle).
For $Q^{++}$ and $Q^{--}$, this ``magnetic mirror'' effect 
increases for large energies $E$. Therefore, a smaller fit region was used for these spectra.
The electron efficiencies of the detectors were verified by analyzing the trigger
functions and found to be 1 above 200 keV.
The proton asymmetry parameter itself varies slightly
($\pm 0.8 \sigma_{\text{stat}}$) with different fit regions and their combinations, 
leading to an additional uncertainty of 0.35\%.

At low energies, there is some HV related background left in the spectra 
(Fig.\ \ref{Fig_CFits}). In
the fit regions, however, they can be well described by fits with a normalizing
factor as the only free parameter, indicating that almost no background is left in this region.
The remaining uncertainty of 0.18\% was determined by using the residuals of the fits. Backscattered
electrons assigned to the wrong detector also distort 
the spectra below 240 keV. At higher energies,
backscattering leads to completely negligible corrections in the order of 0.01\% \cite{Sch07c}.

\begin{table}[b]
\caption{\label{tab_bsame} Correction and errors of the proton asymmetry parameter $C$. 
The extrapolation uncertainty contributes to the statistical and the detector calibration errors. } 
\begin{ruledtabular}
\begin{tabular}{l  rr} 
Effect [\%] & Corr. & Err. \\ \hline
Polarization & 0.30  & 0.10 \\ 
Flip Efficiency &  & 0.10 \\
Data Set: {\it Statistics} &  & 0.44 \\
{\it Fit Region} &  & 0.35 \\
{\it Accidental Coincidences } & $-$0.81 & 0.15 \\ 
{\it Background} & & 0.18 \\
Detector: {\it Gain} &  & 0.38 \\
{\it Offset} &  & 0.82 \\ 
{\it Resolution} &  & 0.12 \\
Spectrometer: {\it Mirror Effect} & 0.01 & 0.06 \\
{\it Edge Effect} & 0.26 & 0.05 \\ 
{\it Grid Effect} & 0.08 & 0.05 \\ \vspace{0.1cm}
Correlations: $A$, $B$, $a$ &  & 0.07 \\ 
Sum & $-$0.16 & 1.11 \\
\end{tabular}
\end{ruledtabular}
\end{table}

Corrections associated with the design of the spectro\-meter, ``Mirror Effect'',
``Edge Effect'', and ``Grid Effect'', are described in \cite{Sch07}. They were
determined by Monte Carlo simulations. For the fit regions of this analysis, all associated
errors and corrections to $C$ are small. The same holds for the influence of the correlation
coefficients $A$, $B$, and $a$ entering via eq.~(\ref{QFunktion}).

Fig.\ \ref{Fig_CFits} shows the fits and integrals for one detector
yielding the result
\begin{equation}\label{eq_result}
C = -0.2377(10)_{\text{stat}}(24)_{\text{syst}}= -0.2377(26).
\end{equation}
A detailed summary of all corrections and uncertainties 
can be found in \mbox{Table \ref{tab_bsame}}.
The second proton detector had a much worse proton efficiency 
due to an inferior carbon foil quality,
leading to an asymmetry result with am more than $3\times$ larger uncertainty. Therefore we
only consider the precise value here, the second serves as cross 
check: Both values agree within their uncertainties.
We also do not consider the result of the earlier measurement \cite{Kre04}
with \mbox{PERKEO II} due to its larger systematic uncertainties.

Our final result for the proton asymmetry parameter, eq.~(\ref{eq_result}), has
an uncertainty of only 1.1\%.
Note that it is correlated with the recently published neutrino asymmetry
parameter $B$ \cite{Sch07} as the analyses
use partly the same data ($Q^{++}$, $Q^{--}$).
Our result agrees with \mbox{$C_{\text{SM}} = -0.2392(4)$}, 
the SM expectation calculated using eq.~(\ref{eq_CandLambda}) and the
world mean value $\lambda=-1.2695(29)$ \cite{PDG06}. It also
fulfills relation (\ref{eq_relat}), but the precision of $C$ has to be improved
for stringent SM cross checks. 
However, for the first time, a proton asymmetry parameter 
can now be included in general parameter space scans of low energetic beta-decays 
\cite{Sev06} to set limits on processes beyond the SM.

\begin{figure}[t!]
\includegraphics*[width=8.5cm]{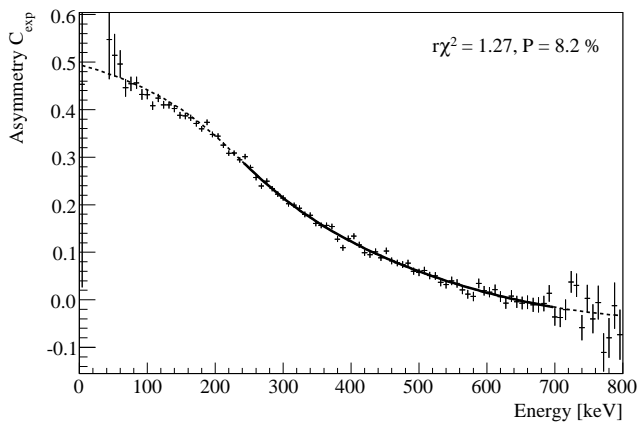}
\caption{\label{Fig_lambda} The experimental proton asymmetry $C_{\text{exp}}(E)$ of proton detector 2 and the fit
in order to obtain $\lambda$. 
Since some systematic effects cancel in the asymmetry, the asymmetry
can be described very well 
over a large energy range. The fit region is indicated by the bold line.}
\end{figure}

We can use eq.~(\ref{eq_CandLambda}) to derive $\lambda_{\text{(int)}}=-1.282(21)$
from the integral proton asymmetry $C$. A more sensitive and direct determination of $\lambda$
from the same data can be performed by a differential analysis of the electron energy $E$
dependent proton asymmetry
\begin{equation}\label{eq_cexp}
  C_{\text{exp}}(E) = \frac{(Q^{++}+Q^{-+}) - (Q^{+-}+Q^{--})}{(Q^{++}+Q^{-+}) + (Q^{+-}+Q^{--})}.
\end{equation}
The dependencies of the $Q^{ij}$ on $E$ have been omitted here. The fit function is
calculated using eq.~(\ref{QFunktion}) and replacing
the coefficients $a$, $A$, and $B$ by their SM expressions in terms of $\lambda$
(as given in, e.g., \cite{Glu95}). In order to avoid systematic effects
due to different proton efficiencies, we again use always the same hemisphere for
proton detection and flip the spin. Consequently, the asymmetry~(\ref{eq_cexp}) contains data 
from different electron detectors with different calibration 
uncertainties. In order to estimate the related systematic effects on $\lambda$, we 
consider the properties of the inferior detector. Again, this uncertainty dominates the
overall error budget. 

On the other hand, systematic effects such as the detector threshold and the 
magnetic mirror effect cancel
in the asymmetry spectrum (\ref{eq_cexp}) allowing to describe the data over a large
energy region (Fig.\ \ref{Fig_lambda}): We can extend the
fit region to up to 700 keV. The lower border is again 240 keV to avoid systematic
effects due to different electron detector thresholds or backscattering. The result
\begin{equation}\label{eq_reslambda}
  \lambda_{\text{(diff)}}=-1.275(6)_{\text{stat}}(15)_{\text{syst}}=-1.275(16)
\end{equation} 
does not depend on the fit region as results for different regions agree within 
$0.2 \sigma_{\text{stat}}$. The overall uncertainty of 1.2\% is dominated by
detector calibration (gain 0.68\%, offset 0.82\%) and statistics (0.50\%). All other
uncertainties are much smaller. This value agrees with the value
$\lambda_{\text{(int)}}$ derived from the integral proton asymmetry
which is an important cross check for the integration and extrapolation procedure.

We recommend the value $\lambda_{\text{(diff)}}$, eq.~(\ref{eq_reslambda}),
as result for $\lambda$ from our experiment as it
is more precise and more direct than $\lambda_{\text{(int)}}$.
It agrees well with the world
mean value $\lambda$ \cite{PDG06} derived from measurements of the beta
asymmetry $A$. It is almost one order of magnitude less precise but has been 
obtained using 
a new method with different systematics: Background is virtually no
issue due to the coincidence condition.

Work was funded by the German Federal Ministry for Research and Education, 
contracts 06HD153I, 06HD187.

\end{document}